\documentclass[aps,prl,twocolumn,showpacs]{revtex4}
\usepackage{amssymb,amsmath}
\usepackage{graphicx}
\usepackage{epsf}
\usepackage{psfig}

\begin{document}

\title{Quantum critical end point of the Kondo volume collapse}
\author{M. Dzero$^1$, M. R. Norman$^2$, I. Paul$^{2,3}$, C. P\'{e}pin$^3$,
and J. Schmalian$^1$}
\affiliation{$^1$Department of Physics and Astronomy, Iowa State University and Ames
Laboratory, Ames, IA 50011\\
$^2$Materials Science Division, Argonne National Laboratory, Argonne, IL
60439 \\
$^3$SPhT, CEA-Saclay, L'Orme des Merisiers, 91191 Gif-sur-Yvette, France}
\date{\today}

\begin{abstract}
The Kondo volume collapse describes valence transitions in $f$-electron
metals, and is characterized by a line of first order transitions in the
pressure-temperature phase plane terminated at critical end points. We
analyze the quantum critical end point, when the lower end point is tuned to 
$T=0$, and determine the specific heat, thermal expansion, and
compressibility. We find that the inclusion of quantum critical fluctuations
leads to a novel bifurcation of the first order phase line. Finally, we show
that critical strain fluctuations can cause \ both, superconductivity \ and
non-Fermi liquid behavior near the critical point.
\end{abstract}

\maketitle

Quantum criticality in heavy fermion materials is predominantly discussed in
the context of magnetic phase transitions\cite{Greg}. In contrast,
instabilities in the charge sector, which are well known to occur in \textrm{%
Ce} and \textrm{Yb} based intermetallics, are usually first order
transitions. In the case of strong first order transitions, like the Kondo
volume collapse (KVC) transition between $\alpha $ and $\gamma $ \textrm{Ce} 
\cite{RPP} or the related behavior in \textrm{YbIn}$\mathrm{Cu}_{4}$ \cite%
{Sarrao}, critical fluctuations are irrelevant. However, theories \cite%
{Allen82, Lavagna82} and experiments \cite{Lawrence84} for the KVC yield a
line of first order phase transitions in the pressure-temperature phase
plane that is terminated at critical points on the high and (sometimes) on
the low temperature end. The location of this lower end point is tunable.
For example, a suppression of the volume collapse transition was achieved by
alloying \textrm{Ce} with \textrm{Th} \cite{Lawrence84} or by applying an
external magnetic field to \textrm{Ce}$_{0.8}$\textrm{La}$_{0.1}$\textrm{Th}$%
_{0.1}$ \cite{Smith}. In the case where the lower end point is tuned to $T=0$%
, a quantum critical end point emerges.

In this paper we investigate the behavior in the vicinity of the quantum
critical end point of the Kondo volume collapse. As shown in Fig.~1, we
demonstrate that the $p-T$ phase transition line is qualitatively changed by
critical fluctuations. In addition, we determine the pressure and
temperature dependence of the heat capacity, the compressibility, and the
thermal expansion, and discuss the possibility of superconductivity and
non-Fermi liquid behavior caused by these critical fluctuations. \textrm{Ce}
and \textrm{Yb} based intermetallic alloys with a moderate high temperature
bulk modulus are good candidates to identify such a quantum critical point
of the Kondo volume collapse, where the critical fluctuations are in a
universality class similar to that of a metamagnetic end point\cite{Millis02}%
. Being in the charge sector, KVC fluctuations may serve as a mechanism for $%
s$-wave superconductivity \cite{Razafim, Miyake99}. Strong retardation
effects of the critical pairing interaction are important and enhance the
superconducting transition temperature as compared to a BCS calculation. Our
predictions for the temperature dependence of the compressibility and the
thermal expansion are unique for the quantum critical end point of the Kondo
volume collapse. Identifying such a behavior in a superconducting material
strongly suggests that the pairing is due to critical strain fluctuations.
Finally, recent advances in the theory of \ quantum phase transitions in
metallic systems\cite{Chubukov03} demonstrate that previous results obtained
for magnetic critical points are unstable against non-analytic corrections
to the collective mode dynamics. Quantum criticality in the charge sector,
as discussed in our paper, is therefore one of the very few cases where
robust predictions for non-Fermi liquid behavior can be made.

\begin{figure}[tbp]
\centerline{\psfig{file=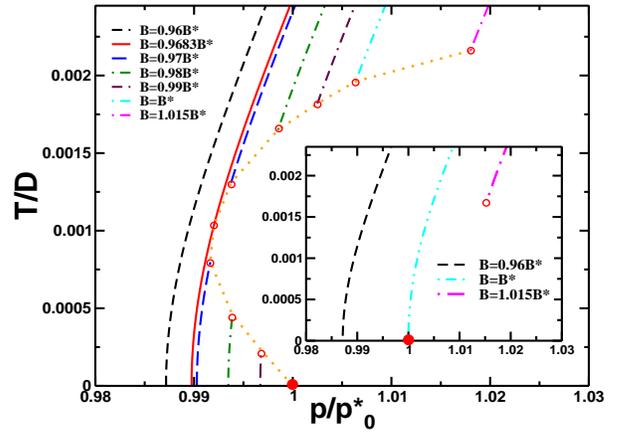,height=6cm,width=8cm,angle=-90}}
\caption{Pressure-temperature phase
diagram for the KVC model for different values of the bare bulk modulus $B$
from the slave boson theory (see text). The main figure includes critical
fluctuations, the inset is the mean field result. Lines represent first
order transitions from a low $T_{K}$ phase, to the left of the line, to a
high $T_{K}$ phase, to the right of the line (this is for Ce alloys, the
reverse would be true for Yb ones). The solid circles and dot  indicate
critical end points for $T>0$ and $T=0$, respectively. The inclusion of
fluctuations causes the first order line to \textquotedblleft bifurcate" for 
$B<B^{\ast }$. The lower line then terminates at a quantum critical end
point for $B=B^{\ast }$.}
\label{fig1}
\end{figure}

A candidate for this behavior is \textrm{Ce}$\mathrm{Cu}_{2}\mathrm{Si}_{2-x}%
\mathrm{Ge}_{x}$ under pressure \cite{Yuan03}. A volume discontinuity at $%
p\simeq 15 \mathrm{GPa}$ was observed in \textrm{Ce}$\mathrm{Cu}_{2}\mathrm{%
Ge}_{2}$ \cite{ssc}. Evidence for a valence instability in \textrm{Ce}$%
\mathrm{Cu}_{2}\mathrm{Si}_{2}$ at pressure $p\simeq 25 \mathrm{kbar}$ was
already given in Ref.~\onlinecite{Bellarbi84}. The increase of the residual
resistivity and the drop of the $T^{2}$ coefficient of the low temperature
resistivity with pressure in \textrm{Ce}$\mathrm{Cu}_{2}\mathrm{Si}_{2}$ and 
\textrm{Ce}$\mathrm{Cu}_{2}\mathrm{Ge}_{2}$ have been interpreted in terms
of critical valence fluctuations \cite{Miyake99, Miyake02}. Also, the
ability to separate two superconducting regimes by varying pressure in 
\textrm{Ce}$\mathrm{Cu}_{2}\mathrm{Si}_{1.2}\mathrm{Ge}_{0.8}$ \cite{Yuan03}
supports the point of view that two different mechanisms are at work, one
related to the magnetic critical point and the other to the valence
fluctuation one \cite{Miyake99}.

The order parameter at a volume collapse transition between two isotructural
states is the trace 
\begin{equation}
\varepsilon \equiv \mathrm{tr}\widehat{\varepsilon }=\frac{V-V_{0}}{V_{0}}
\end{equation}%
of the strain tensor $\widehat{\varepsilon }$, where $V_{0}$ is a reference
volume. Our choice for $V_{0}$ as the volume at the quantum critical end
point will be discussed below. The mean field theory for the KVC transition 
\cite{Allen82, Lavagna82} starts from the Gibbs free energy 
\begin{equation}
G\left( p,T\right) =F_{\mathrm{K}}\left( \varepsilon ,T\right)
+pV_{0}\varepsilon +\frac{1}{2}BV_{0}\varepsilon ^{2}  \label{AM}
\end{equation}%
where $B$ is the bare (high temperature) bulk modulus that describes the
elastic properties in the absence of the Kondo effect at the volume $V_{0}$.
In a cubic system $B=\frac{1}{3}\left( c_{11}^{0}+2c_{12}^{0}\right) $ with
bare elastic constants $c_{11}^{0}$ and $c_{12}^{0}$, $p$ is the pressure,
and $F_{\mathrm{K}}\left( \varepsilon ,T\right) $ the Kondo contribution to
the free energy. For $T\ll T_{K}$, $F_{\mathrm{K}}\left( T,V\right) \simeq
-V_{0}\left( T_{K}+a\frac{T^{2}}{T_{K}}\right) $ with Kondo temperature $%
T_{K}\simeq De^{-1/J_{K}}$ ($D$ is of order the bandwidth of the conduction
electrons) and $a$ of order unity. Due to the exponential variation of $%
T_{K} $ with $J_{K}$, moderate variations of the Kondo coupling $J_{K}$ with
volume cause a strong volume dependence of the Kondo temperature. This, in
turn, yields a non-linear equation of state $p\left( V\right) =-\left. \frac{%
\partial F}{\partial V}\right\vert _{T}=-\frac{1}{V_{0}}\left. \frac{%
\partial F}{\partial \varepsilon }\right\vert _{T}$. It is easy to verify
that with appropriate values of $B$, even a linear volume dependence of $%
J_{K}$ yields $p-V$ isotherms similar to those of the van der Waals theory
of the liquid-gas transition \cite{Allen82,Lavagna82}, with a first order
transition where a discontinuous volume change occurs. For sufficiently
large $B$, the line of first order transitions is terminated at a lower
critical end point. For $T\gg T_{K}$, $\ F_{\mathrm{K}}$ is dominated by a
saturating spin entropy, resulting in a termination of the line of first
order transitions at a high $T$ critical end point \cite{Dzero}.

In order to go beyond mean field theory and to analyze the quantum critical
fluctuations of the KVC transition, we consider $\varepsilon \left( \mathbf{%
x,}\tau \right) $ as a space and time dependent fluctuating strain with a
field theory governed by the action 
\begin{equation}
S\left[ \varepsilon \right] =S_{\mathrm{fl}}\left[ \varepsilon \right] +\int
d^{d}x\int_{0}^{\beta }d\tau \left( \Phi \left( \varepsilon \left( \mathbf{x}%
,\tau \right) \right) +p\varepsilon \left( \mathbf{x},\tau \right) \right) .
\label{stot}
\end{equation}
Here, $S_{\mathrm{fl}}\left[ \varepsilon \right] $ describes strain
fluctuations that are nonlocal in time and space while the static potential $%
\Phi \left( \varepsilon \right)$ includes the nonlinear effects of the
theory. The external pressure $p$ occurs in the last term such that the
Gibbs free energy follows as $G\left( p\right) =-T\log \int d\varepsilon
\exp \left( -S\right)$. We first derive Eq.~\ref{stot} from a microscopic
model. Then we discuss the physical consequences that follow from the
analysis of $S\left[ \varepsilon \right]$.

To describe the Kondo lattice physics of the system, we start from the
infinite $U$ Anderson lattice model 
\begin{equation}
H=H_{0}+\sum_{i\sigma }\left( \epsilon _{f}^{0}f_{i\sigma }^{\dagger
}f_{i\sigma }+t\left( f_{i\sigma }^{\dagger }c_{i\sigma }+h.c.\right)
\right) 
\end{equation}%
where $H_{0}=\sum_{\mathbf{k}\sigma }\epsilon _{\mathbf{k}}c_{\mathbf{k}%
\sigma }^{\dagger }c_{\mathbf{k}\sigma }$ is the Hamiltonian of the
conduction electrons, $\epsilon _{f}^{0}$ the bare $f$ electron energy, and $%
t$ the hybridization. The model describes strongly correlated electrons
because of the infinite Coulomb repulsion of the $f$ electrons, leading to $%
n_{fi}=\sum_{\sigma =1}^{N}f_{i\sigma }^{\dagger }f_{i\sigma }\leq 1$ ($N$
is the degeneracy of the $f$ electrons).

In a situation where the value of the hybridization depends sensitively on
the volume of the system, i.e. on the local strain $\varepsilon _{i}$, we
replace $H\rightarrow H+H_{\mathrm{c}}$ with 
\begin{equation}
H_{\mathrm{c}}=\gamma t\sum_{i\sigma }\varepsilon _{i}\left( f_{i\sigma
}^{\dagger }c_{i\sigma }+h.c.\right) \ +\frac{B}{2}\frac{V_{0}}{N_{0}}%
\sum_{i}\varepsilon _{i}^{2}  \label{Hc}
\end{equation}%
where $\gamma $ is the coefficient of the assumed linear volume dependence
of the hybridization. $B$ is, as in Eq.~\ref{AM}, the bare bulk modulus and $%
N_{0}$ the number of unit cells in the system. In Eq.~\ref{Hc} we assumed
that $\varepsilon $ only couples to the hybridization. Additional couplings
to the conduction band and $f$ level energies also occur, but will not
qualitatively change the conclusions of our paper. They will be ignored in
what follows.

In order to obtain a quantitative insight into the low temperature behavior
of the Kondo lattice, we use a slave boson mean field approach \cite%
{Millis87} to enforce the constraint $n_{fi}\leq 1$. The main results of
this paper do not depend on the details of this method, though. The result
of the slave boson mean field theory is the free energy $\Phi \left(
\varepsilon \right) $ per volume $V_{0}$ as a function of strain. Within
mean field theory $\Phi \left( \varepsilon \right) =F_{K}/V_{0}+\frac{1}{2}%
B\varepsilon ^{2}$ of Eq.~\ref{AM} and results for the volume collapse
transition are in full agreement with Refs.~\onlinecite{Allen82} and %
\onlinecite{Lavagna82}.

\begin{figure}[tbp]
\centerline{\psfig{file=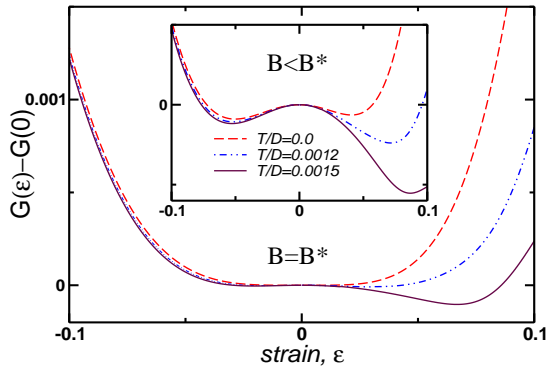,height=6cm,width=8cm,angle=-90}}
\caption{Slave boson mean field results
for the strain dependence of the Gibbs free energy. A double minima
indicates the presence of a first order transition. For $B=B^{\ast }$ and $%
T=0$, the double minima collapse to a single one, indicating the presence of
a quantum critical end point. For $B<B^{\ast }$ (inset), the transition is
always first order.}
\label{fig2}
\end{figure}

In Fig.~2 we show the results for the pressure-strain dependence of the
Gibbs free energy $G(\varepsilon )=\Phi (\varepsilon )+p\varepsilon $ for
different values of the bare bulk modulus as obtained from the slave boson
mean field calculation. Close to the transition and at low $T$ the potential
has the form 
\begin{equation}
\Phi \left( \varepsilon \right) =\Phi _{0}-p^{\ast }\varepsilon +\frac{b}{2}%
\varepsilon ^{2}-\frac{v}{3}\varepsilon ^{3}+\frac{u}{4}\varepsilon ^{4}
\label{pot3}
\end{equation}%
Here, $\Phi _{0}$ is the $\varepsilon $ independent part of $\Phi
(\varepsilon )$ and $p^{\ast }\left( T=0\right) $ the pressure at the
critical point. The coefficients $p^{\ast }$, $b$, $v$ and $u$ can be
obtained from the slave boson theory by differentiating the free energy with
respect to the strain. They are temperature dependent due to the temperature
dependence of the spin entropy, and thus at low temperatures are functions
of $\frac{T^{2}}{T_{0}}$, where $T_{0}$ is of order the Kondo temperature.
Note that $b$ also has a term determined by the bare bulk modulus $B$
(Eq.~2). The lower end point of the KVC transition is located at $T=0$, if $%
\frac{\partial p}{\partial \varepsilon }=\frac{\partial ^{2}p}{\partial
\varepsilon ^{2}}=0$. This is achieved by varying $B$ to the value $B^{\ast
} $ and the pressure to $p^{\ast }$. In order to have this transition take
place at $\varepsilon =0$, we choose the so far arbitrary reference volume $%
V_{0}$ such that $\Phi \left( \varepsilon \right) $ has no contribution of
order $\varepsilon ^{3}$ at $T=0$. This implies that $v=\varkappa {T}%
^{2}/T_{0}$ with $\varkappa >0$ and that $p^{\ast }$ now depends on $B$.

The dynamic and nonlocal part $S_{\mathrm{fl}}$ of the action can also be
derived within the slave boson approach. Integrating out the fermionic
degrees of freedom, 
\begin{equation}
S_{\mathrm{fl}}=\frac{1}{2}\int_{q}\left( \Pi \left( q\right) -\Pi \left(
0\right) \right) \left\vert \varepsilon \left( q\right) \right\vert ^{2}
\end{equation}%
where $\int_{q}...=T\sum_{n}\int \frac{d^{d}\mathbf{q}}{\left( 2\pi \right)
^{d}}...$ with $q=\left( \mathbf{q,}\omega _{n}\right) $ and $\Pi \left(
q\right) =-N\gamma ^{2}t^{2}\int_{q^{\prime }}\left( G_{q^{\prime
}}^{cc}G_{q^{\prime }+q}^{ff}+G_{q^{\prime }}^{cf}G_{q^{\prime
}+q}^{cf}\right) $. The polarization bubble $\Pi \left( {\mathbf{q}},\omega
\right) $ determines the dynamics of the strain field and is determined by
the single particle propagators of the fermions within the slave boson
calculation, i.e. $G_{\mathbf{q}}^{cc}\left( \tau \right) =-\left\langle
T_{\tau }c_{\mathbf{q}}\left( \tau \right) c_{\mathbf{q}}^{\dagger }\left(
0\right) \right\rangle $, etc. The evaluation of this expression for small $%
\mathbf{q}$ and $\omega $ (but with $v_{F}\left\vert \mathbf{q}\right\vert
\gg \omega $) yields\cite{parameters} 
\begin{equation}
S_{\mathrm{fl}}=\frac{1}{2}\int_{q}\left( \Gamma \frac{\left\vert \omega
_{n}\right\vert }{v_{F}\left\vert \mathbf{q}\right\vert }+\left( \frac{%
\alpha \mathbf{q}}{2k_{F}}\right) ^{2}\right) \left\vert \varepsilon \left(
q\right) \right\vert ^{2}  \label{sfl}
\end{equation}%
As long as the system is in a heavy fermi liquid regime, the generic $\omega 
$ and $\mathbf{q}$-dependence of $S_{\mathrm{fl}}$ in Eq.~\ref{sfl} is
independent of the details of the slave boson approach. The latter does
however allow one to determine the prefactors $\Gamma $ and $\alpha $ in
terms of microscopic quantities\cite{Millis87,parameters}. The model, Eqs.~%
\ref{stot} and \ref{sfl}, is essentially the same as the one used by Millis 
\emph{et al.} \cite{Millis02} in the context of metamagnetic quantum
criticality in metals and numerous analogies exist.

From the effective action, Eqs.~\ref{stot}, \ref{pot3} and \ref{sfl}, we
obtain the equation of state including fluctuation corrections and determine
the location of the volume collapse transition. It follows that 
\begin{equation}
p=p^{\ast }-\left( b+\ u\left\langle \varepsilon ^{2}\right\rangle \right)
\varepsilon +v\varepsilon ^{2}-u\varepsilon ^{3}  \label{eos1}
\end{equation}%
with $\left\langle \varepsilon ^{2}\right\rangle _{B=B^{\ast
}}=\int_{q}\left( \frac{\Gamma \left\vert \omega _{n}\right\vert }{%
v_{F}\left\vert \mathbf{q}\right\vert }+\left( \frac{\alpha \mathbf{q}}{%
2k_{F}}\right) ^{2}\right) ^{-1}$. There are two distinct sources of
temperature variation in Eq.~\ref{eos1}. First, due to the temperature
dependence of the mean field entropy, it holds for $T\ll T_{0}$: $p^{\ast
}\left( T\right) =p_{0}^{\ast }+\zeta ^{\prime }\frac{T^{2}}{T_{0}}$, $%
b\left( T\right) =b_{0}-\zeta \frac{T^{2}}{T_{0}}$, etc., where $\zeta
,\zeta ^{\prime }$ are positive constants. Second, the order parameter
fluctuations are temperature dependent: $\left\langle \varepsilon
^{2}\right\rangle _{T}=\left\langle \varepsilon ^{2}\right\rangle _{T=0}+%
\tilde{\zeta}T^{\frac{d+1}{3}}$ with $\tilde{\zeta}\simeq \frac{12.15\Gamma
^{1/3}}{\alpha ^{8/3}(v_{F}k_{F})^{1/3}}>0$ for $d=3$. For $d<5$, the latter
is the dominant effect (fluctuation corrections due to the cubic term $%
\propto v$ are subleading). 

We first ignore the $T$ dependence of $\left\langle \varepsilon
^{2}\right\rangle $, formally corresponding to $d>5$, and analyze the mean
field $p-T$ phase diagram. The critical end point is located at $T=0$ if $%
p=p_{0}^{\ast }$ and $B-B^{\ast }\equiv b_{0}+\ u\left\langle \varepsilon
^{2}\right\rangle _{T=0}=0$. Close to this transition at $T=0$, $\varepsilon
\left( B=B^{\ast },p\right) \propto \left( {p_{0}^{\ast }-p}\right) ^{1/3}$
and $\varepsilon \left( B,p=p_{0}^{\ast }\right) \propto (B^{\ast }-B)^{1/2}$%
. At finite $T$ we obtain a first order transition if $B<B^{\ast }+\zeta 
\frac{T^{2}}{T_{0}}$. For $B>B^{\ast }+\zeta \frac{T^{2}}{T_{0}}$, no
transition takes place. The zero temperature transition is first order for $%
B<B^{\ast }$, second order for $B=B^{\ast }$, and absent for $B>B^{\ast }$.
For the following set of parameters: $t=0.5\mathrm{eV}$, $D=1$\textrm{eV}, $%
\varepsilon _{f}^{0}=-2.34\mathrm{eV}$, (yielding $T_{K}\simeq 0.012D$ and $%
n_{f}=0.87$), $B^{\ast }=254$\textrm{kbar}, $p_{0}^{\ast }=-4.7$kbar and $%
\gamma =2.14$, the corresponding $p-T$ curves are shown in the inset of
Fig.~1. The value for $\gamma $ is chosen to yield results\ that
quantitatively reproduce the mean field theory of Ref.\cite{Allen82}.

The phase diagram changes qualitatively if we include the temperature
dependence of the critical fluctuations of the order parameter $\left\langle
\varepsilon ^{2}\right\rangle _{T}$, which is dominant at low temperatures
for $d<5$. Besides entering with a different power than the mean field
terms, the key new aspect of this fluctuation effect is that it leads to an
increase, as opposed to a decrease, of the $b\left( T\right) +\
u\left\langle \varepsilon ^{2}\right\rangle $ term in Eq.~\ref{eos1} as $T$
increases. While the mean field entropy softens the bulk modulus, critical
fluctuations harden it. Thus, if the critical point is located at $T=0$,
increasing temperature will suppress the transition rather than making it go
first order. This effect is demonstrated in Fig.~1, where we show the
fluctuation corrected transition lines for different bulk moduli. The
competition between the fluctuation term (which dominates at the lowest
temperatures) and the mean field terms (which dominate at higher
temperature) leads to the development of a gap in the first order line at a $%
B>0.9683B^{\ast }$. With increasing $B$, the first order line below this gap
eventually collapses to a quantum critical point at $B=B^{\ast }$, and then
disappears for larger $B$. Thus, the critical end point becomes isolated in
the $p-T$ phase plane. For the parameters we used, the gap in the first
order transition $\simeq 0.002D\simeq 20\mathrm{K}$ should be  observable.

Next, we use the equation of state Eq.~\ref{eos1} close to the quantum
critical end point to determine the pressure and temperature dependence of $%
\varepsilon $, the compressibility $\kappa $ and the thermal expansion $%
\beta $. The analysis of Eq.~\ref{eos1} yields 
\begin{eqnarray}
\varepsilon  &=&\frac{\Delta V}{V}\propto T^{\phi },~\kappa =-\frac{1}{V}%
\frac{\partial V}{\partial p}\propto T^{-\gamma }  \notag \\
\beta _{T} &=&\frac{1}{V}\frac{\partial V}{\partial T}\propto T^{-\theta }
\end{eqnarray}%
where $\theta =1-\phi $. We obtain the exponents $\phi =\frac{2}{3}$, $%
\gamma =\frac{4}{3}$ and $\theta =\frac{1}{3}$ for $d\geq 3$ and $\phi =%
\frac{5-d}{3}$, $\gamma =\frac{d+1}{3}$ and $\theta =\frac{d-2}{3}$ for $d<3$%
. For $d<$ $3$ the $T$-dependence of the strain is governed by the
fluctuation corrections in the equation of state, while for $d>3$, the mean
field $T$ dependence $\frac{T^{2}}{T_{0}}$ in $\Phi \left( \varepsilon
\right) $ of Eq.~\ref{pot3} dominates. For $d=3$ both terms yield exactly
the same $T$ dependence. The strain fluctuations also give rise to a
singular heat capacity $\frac{C_{V}}{T}$ $\propto -\log T$ for $d=3$ and $%
\frac{C_{V}}{T}$ $\propto T^{\frac{d-3}{3}}$ for $d<3$. For the physically
relevant heat capacity at constant pressure $C_{p}=C_{V}+V\kappa {T}\left( {%
\partial {p}}/{\partial {T}}\right) _{V}^{2}$ we find $C_{p}-C_{V}\propto {T}%
^{3-\gamma }$, i.e. the dominant contribution to $C_{p}$ comes from $C_{V}$.
Pressure tuning at $T=0$ yields a diverging compressibility: $\kappa \propto
\left\vert p-p_{0}^{\ast }\right\vert ^{-2/3}$. Critical fluctuations in the
charge sector furthermore yield non-Fermi liquid behavior of the single
particle self energy: $\Sigma _{k_{F}}\left( \omega \right) \propto \omega
\left\vert \omega \right\vert ^{-\frac{3-d}{3}}$ for $d<3$ and $\Sigma
_{k_{F}}\left( \omega \right) \propto \omega \log \left\vert \omega
\right\vert $ for $d=3$. 

Finally we comment on the possibility of superconductivity caused by the
interaction of electrons with these critical fluctuations. Strain
fluctuations couple to electrons just like phonons or other bosonic charge
excitations, leading to an attractive interaction in the s-wave pairing
channel. Superconductivity as caused by valence fluctuations was discussed
in Refs.~\onlinecite{Razafim} and \onlinecite{Miyake99}. In those
treatments, retardation effects of the pairing interaction enter the theory
solely via the upper cut off, $\omega _{0}$, of the theory. In the weak
coupling limit this leads to the well known BCS-formula for the transition
temperature $T_{c}^{BCS}\simeq \omega _{0}e^{-\frac{1}{\lambda _{p}}}$ where 
$\lambda _{p}\propto \frac{\gamma ^{2}t^{2}}{D^{2}}$ is the dimensional
coupling constant of the pairing interaction. However, in the vicinity of
the critical end point, the strain fluctuations serve as pairing boson which
becomes massless. As shown in Ref.~\onlinecite{Son99,Chubukov05} retardation
effects are more subtle in the case of pairing due to a massless boson
governed by Eq.~\ref{sfl}. For a given value of $\lambda _{p}$, the
transition temperature is enhanced compared to the BCS result and it follows
that 
\begin{equation}
T_{c}\simeq \omega _{0}e^{-\frac{\pi }{2\sqrt{\lambda _{p}}}}.
\end{equation}%
Thus, the critical fluctuations discussed in this paper are good candidates
for bosons that cause $s$-wave superconductivity.

In summary, we have analyzed the critical behavior in the vicinity of a $T=0$
end point for the Kondo volume collapse transition. We find diverging
specific heat and thermal expansion coefficients, and a diverging
compressibility. We also find that critical fluctuations break apart the
line of first order transitions, leading to a novel \textquotedblleft
isolation" of the critical end point in the pressure-temperature phase
plane. Finally, we discussed the implications of our results for the
existence of superconductivity near a KVC quantum critical end point.

The authors thank the hospitality of the KITP where this work was initiated
and A. V. Chubukov and P. Coleman for useful duscussions. This work was
supported by the U.S. Dept. of Energy, Office of Science, under Contracts
No.~W-31-109-ENG-38 (ANL) and W-7405-ENG-82 (Ames), and in part by the
National Science Foundation under Grant No.~PHY99-07949.


\begin{thebibliography}{99}
\bibitem{Greg} G. R. Stewart, Rev. Mod. Phys. \textbf{73}, 797 (2001).

\bibitem{RPP} J. M. Lawrence, P. S. Riseborough, R. D. Parks, Rep. Prog.
Phys. \textbf{44}, 1 (1981).

\bibitem{Sarrao} J. L. Sarrao, Physica B \textbf{259-261}, 128 (1999).

\bibitem{Allen82} J. W. Allen and R. M. Martin, Phys. Rev. Lett. \textbf{49}%
, 1106 (1982).

\bibitem{Lavagna82} M. Lavagna, C. Lacroix, M. Cyrot, Phys. Lett. \textbf{90}%
A, 210 (1982) and J. Phys. F \textbf{13}, 1007 (1983).

\bibitem{Lawrence84} J. M. Lawrence \textit{et al.}, 
Phys. Rev. B \textbf{29}, 4017 (1984).

\bibitem{Smith} F Drymiotis \textit{et al.}, 
J. Phys. Condens. Matter \textbf{17}, L77 (2005).

\bibitem{Millis02} A. J. Millis, A. J. Schofield, G. G. Lonzarich, S. A.
Grigera, Phys. Rev. Lett. \textbf{88}, 217204 (2002).

\bibitem{Razafim} H. Razafimandimby, P. Fulde, J. Keller, Zeit. Phys. 
\textbf{54}, 111 (1984).

\bibitem{Miyake99} K. Miyake, O. Narikiyo, Y. Onishi, Physica B \textbf{%
259-261}, 676 (1999).


\bibitem{Chubukov03} A. V. Chubukov and D. L. Maslov, Phys. Rev. B \textbf{68%
}, 155113 (2003); D. Belitz, T.R. Kirkpatrick, and T. Vojta, Phys. Rev. B 
\textbf{55}, 9452 (1997).

\bibitem{Yuan03} H. Q. Yuan \textit{et al.}, 
Science \textbf{302}, 2104 (2003).


\bibitem{ssc} A. Onodera \textit{et al.}, Solid State Comm. \textbf{123},
113 (2002).

\bibitem{Bellarbi84} B. Bellarbi \textit{et al.}, 
Phys. Rev. B \textbf{30}, 1182 (1984).

\bibitem{Miyake02} K. Miyake and M. Maebashi, J. Phys. Soc. Jpn. \textbf{71}%
, 3955 (2002).



\bibitem{Dzero} M. Dzero, L. P. Gor'kov, A. K. Zvezdin, J. Phys. Condens.
Matter \textbf{12}, L711 (2000).

\bibitem{Millis87} P. Coleman, Phys. Rev. B \textbf{29}, 3035 (1984); A. J.
Millis and P. A. Lee, Phys. Rev. B \textbf{35}, 3394 (1987).

\bibitem{parameters} The slave boson approach yields, $\Gamma =\frac{N\pi
\rho _{F}\left( 1-n_{f}\right) m^{\ast 2}\gamma ^{2}t^{2}}{2m^{2}\varepsilon
_{f}^{02}}$and $\alpha ^{2}=\frac{N\rho _{F}\left( 1-n_{f}\right) \gamma
^{2}t^{2}}{3\epsilon _{f}^{0}D}$ . $m$ is the bare mass of the conduction
electrons, $\rho _{F}=mk_{F}/2\pi ^{2}$ is the density of states at the
Fermi level, $m^{\ast }$ and $\left( k_{F}/k_{0F}\right) ^{2}=1+\left(
1-n_{f}\right) t^{2}/\left( \epsilon _{f}^{0}D\right) $ are the renormalized
mass and Fermi momentum of the heavy quasiparticles below $T_{K}$.






\bibitem{Son99} D. T. Son, Phys. Rev. D \textbf{59}, 094019 (1999).

\bibitem{Chubukov05} A. V. Chubukov and J. Schmalian, Phys. Rev. B \textbf{72%
}, 174520 (2005).
\end{thebibliography}
\end{document}